\documentclass[preprint, showpacs, preprintnumbers, amsmath,amssymb,graphicx]{revtex4}

\usepackage{graphicx}
\usepackage{dcolumn}
\usepackage{tensor}
\usepackage{fontenc}

\begin{document}

\title{ The modified wave function of test particle approaching holographic screen from entropy force  }
\author{Bin Liu}
\author{Yun-Chuan Dai}
\author{Xian-Ru Hu}
\author{Jian-Bo Deng}
\email{dengjb@lzu.edu.cn} \affiliation{
Institute of Theoretical Physics, Lanzhou University \\
Lanzhou 730000, People's Republic of China}

\date{\today}

\begin{abstract}

 In this note we generalize entropy based on the quantum mechanical probability density distribution. Motivated by J. Munkhammar
 and the uncertainty of entropy we modified the origin wave function of the test particle. The corrected one acting
 on the quantum particle is subject to the uncertainty principle. Considering the uncertainty relation, the corrected probability
 of the particle for measurement on holographic screen has been proposed. We also derivate the speed of information transfer
 described by wave function. Our quantum approach to entropy stress the information in a physical system is directly associated
 with its quantum mechanical entropy defined by the equality of the partition function.


\end{abstract}
\maketitle

\section{introduction}

In a recent paper, Verlinde \cite{f1} has conjectured that the origin of Newtonian gravity and second law of dynamics might be entropic in nature. This 
theory is contributed by Jacobson¡¯s approach \cite{f2}and subsequent work by Padmanabhan \cite{f3}\cite{f4}\cite{f5}. According to this theory, the 
spacetime can be described as an information device made of holographic surfaces on which the information about the physical systems can be stored. 
The information on the screens is described by the information entropy and, as the black hole entropy; it is encoded in a number of bits proportional to 
the area of the screen \cite{f6}. The total energy of the degrees of freedom satisfies the equipartition theorem. Also, it is postulated that the 
information entropy is maximized by the entropic forces that act along the holographic directions and are defined by gradients of the entropy.\\
\indent His paper \cite{f1}attracted quite some attention and several papers from various fields of theoretical physics. The Friedmann equations and the modified Friedmann equations for Friedmann-Robertson-Walker universe were derived utilizing the holographic principle and the equipartition rule of
energy \cite{f7}\cite{f8}\cite{f9}\cite{f10}\cite{f11}. The Newtonian gravity in loop quantum gravity was given by entropic force in ref.\cite{f12}.
In ref.\cite{f13}the Coulomb force was regarded as an entropic force. ref.\cite{f14}showed that the holographic dark energy can be derived from the
entropic force formula. J. Makela pointed out in \cite{f15}that Verlinde's entropic force is actually the consequence of a specific microscopic model of 
spacetime. The similar ideas were also applied to the construction of holographic actions from black hole entropy \cite{f16}while ref.\cite{f17}showed
that gravity has a quantum informational origin. In \cite{f18}, a modified entropic force in the Debye model was presented. Another important discovery
was made by Joakim Munkhammar more recently \cite{f19}, which indicates a deep connection between the quantum mechanical probability density 
distribution and entropy. For other relevant works to entropic force, we refer to e.g.\cite{f20}\cite{f21}\cite{f22}\cite{f30}\cite{f31}\cite{f32}
and references therein.\\
\indent In this paper follow the idea which shows the connection between quantum mechanics and entropic force, we study the wave function $\psi$
of the particle with mass $m$ approaching holographic screen. First, entropy based on the quantum mechanical probability density distribution would be introduced. Second, consider the uncertainty of the separation$\triangle x$, we obtain a modified wave function $\psi$ of the particle and give a profound physical meaning.Third, we derivate the speed of information transfer described by wave function. Finally, the paper ends with a brief summary. In this paper, we set the units of $c=G=\hbar=k_B=1$.\\

\section{ENTROPY BASED ON THE QUANTUM MECHANICAL PROBABILITY DENSITY DISTRIBUTION}\label{SecB}

\indent To make our narration clear, we first review the model of quantum mechanical holographic entropy. Then we apply it to a modified wave function of a test particle. In Verlinde's view the information associated with positions and movements, mass of matter in space is displayed to us on a surface
called holographic principle \cite{f23}\cite{f24}\cite{f25}. Information is stored in discrete bits on the screen and since the number of bits is limited
we get holographic effects. This means that if there is more information on the inside than the amount of information accessible on the screen then
information will be hidden from us as we observe the dynamics. The connection between entropy and information is that the change of information $I$
is the negative change of entropy $S$\\
\begin{equation}\label{eq1}
 \triangle I=-\triangle S.
\end{equation}
\indent Considering a particle with mass $m$ approaching a small piece of holographic screen from the side, Verlinde concluded that the entropy
associated with this process should be Bekenstein entropy:
\begin{equation}\label{eq2}
\triangle S=2 \pi k_B \frac{mc}{\hbar} \triangle x.
\end{equation}
\indent Although the dynamics on the screen is governed by some unknown rules, in ref.\cite{f19}J. Munkhammar proposed quantum entropy which is
used as a generalization of the Bekenstein entropy. In Feynman's approach \cite{f26}\cite{f27}any quantum mechanical system is the sum of all complex amplitudes relating to particles paths from one point to another
\begin{equation}\label{eq3}
\psi = R e^{i\frac{A}{\hbar}}= \sum_n e^{i\frac{A_n}{\hbar}}.
\end{equation}
Where $R^2 = \psi \psi^\dagger = |\psi|^2,$ is the probability density distribution. The probability density for a particle is defined as:
\begin{equation}\label{eq4}
\rho = \psi \psi^\dagger = |\psi|^2.
\end{equation}
When integrating (\ref{eq4}) one could get the probability of a state in a particular domain of space. Here we assume that the integration be over the volume $V_S$ inside a given holographic screen gives unity:
\begin{equation}\label{eq5}
\int_{V_\mathcal{S}} |\psi|^2 dV = 1.
\end{equation}
This excludes the possibility of the particle being outside the screen. Via the Feynman approach (\ref{eq3}) to quantum mechanics
we can conclude that $|\psi|^2$ is related to a sum of states of a quantum system. In light of this we suggest that the probability density
$|\psi|^2$ is in fact related to a partition function $Z$ for different possible states
\begin{equation}\label{eq6}
Z = \frac{1}{|\psi|^2}.
\end{equation}
Note that it is defined by the equality of the partition function with the inverse probability density distribution. Furthermore
we can construct entropy:
\begin{equation}\label{eq7}
S = k_B ln(Z) = - 2 k_B ln|\psi|.
\end{equation}
This relation (\ref{eq7}) between $|\psi|$ and $S$ is used as a generalization of the Bekenstein entropy throughout
ref.\cite{f19}. If we take $\psi$ into special form which is the solution of Klein-Gordon equation:
\begin{equation}\label{eq8}
\psi(x) = A e^{-\frac{mc}{\hbar}x}.
\end{equation}
 Substitute this into eq.(\ref{eq7}) we get:
\begin{equation}\label{eq9}
S = - 2 k_B \Big(-\frac{mc}{\hbar}x + ln(A)\Big).
\end{equation}
Now if we return to (\ref{eq9}) and look at the difference in entropy $\Delta S = S_1-S_2$ from the difference in
$\Delta x = x_1-x_2$ we get:
\begin{equation}\label{eq10}
S_1 - S_2 = \Delta S = 2k_B \frac{mc}{\hbar} x_1 - ln(A) - (2k_B\frac{mc}{\hbar} (x_2) - ln(A))=2k_B \frac{mc}{\hbar}\Delta x.
\end{equation}
which is equivalent to the entropy used in Verlinde's approach up to a factor of $\pi$.
In this framework J. Munkhammar has got the expression for Bekenstein's entropy in the situation of a stationary particle. In the next section motivated by ref.\cite{f19} we proposed a modified wave function $\psi$  of the particle by introducing the uncertainty of the entropy and clearer physical meaning would be showed.

\section{THE MODIFIED WAVE FUNCTION OF THE PARTICLE}\label{SecC}
\indent In light of Verlinde's discovery that entropy might be the source of gravity, Ted Jacobsson stated that a quantization of general relativity is physically as absurd as the quantization of for example the wave equation for sound in air \cite{f2}. In a similar fashion we shall not quantize holographic screen here, but rather construct the mechanism of the particle. Follow the key points of above discussion, we calculate the wave function of a single particle at rest to see how the entropy works. So with eq.(\ref{eq7}), we obtain the real part of wave function
\begin{equation}\label{eq11}
|\psi(x)| =Ae^{-\frac{S}{2k_B}}.
\end{equation}
where $A$ is a normalization constant and $x$ is the radius outwards from the classical position of the particle. It should be stressed here that through eq.(\ref{eq11}) the single particle's wave function could be described by Bekenstein's entropy, which means the information stored on the screen. This relation between wave function and entropy might be a powerful tool which provides new insights into the quantum nature.\\
\indent More explicitly, in \cite{f28}the authors have considered the possibility of having an uncertainty in the entropy in the holographic screen
originating from the fact that there are inherent quantum uncertainties in the position and momentum of a particle. This comes primarily because the
information (or entropy) associated with the screen depends linearly on the distance of the test particle from the screen. Keeping this in mind, in
\cite{f28}, the relation of (\ref{eq2}) has been generalized to:\\
\begin{equation}\label{eq12}
\delta S=2 \pi k_B \left (\frac{|\delta x|}{l_p} + \frac{|\delta p|}{mc}\right).
\end{equation}
Here the variations $\delta x$ and $\delta p$ are to considered as quantum uncertainties out of holographic screen, obeying the Heisenberg uncertainty relation:
\begin{equation}\label{eq13}
\delta x \delta p \geqslant \frac{\hbar}{2}.
\end{equation}
Indeed in the classical case $\delta x$ reduces to the separation $\triangle x$ between the screen and the
particle. If $\delta p=0$ as the Heisenberg uncertainty relation does not apply, eq.(\ref{eq12}) reduces to the Verlinde's formula
(\ref{eq2}). Otherwise, one can replace $\delta p$ in (\ref{eq12}) by $\delta p=\frac{\hbar}{2 \delta x}$ to obtain a quantum corrected
$\delta S$:
\begin{equation}\label{eq14}
\delta S=2 \pi k_B \left (\left|\frac{\delta x}{l_p}\right| + \left|\frac{l_p}{2 \delta x}\right|\right).
\end{equation}
When $\delta x \rightarrow 0$, $\delta S \rightarrow \infty$, so we can not determine both $S$ and displacement $x$ simultaneously,
which agrees with the result from ref.\cite{f28}:
\begin{equation}\label{eq15}
\delta F \delta x^2 \geqslant \frac{\hbar}{2m} \left(\frac{\hbar a}{c^2} - p \right).
\end{equation}
General uncertainty relation (\ref{eq15}) tells us when $\delta x \rightarrow 0$, $\delta F \rightarrow \infty$, so it is the same story that
we can not determine both force $F$ and displacement $x$ simultaneously either. Assumption (\ref{eq2}) from Verlinde implies when
observing a physical system the entropy of the system will change because information is obtained on it. In practice this suggests the increase of
entropy on the holographic screen is produced by the information lost from the particle when it is approaching the screen, which generates the
entropy force. So it becomes quite understandable that when $\delta S$ is very large $\delta F$ becomes very large too. If $\delta x$
is back to the separation $\triangle x$ , we derive the following equation by integrating (\ref{eq14})
\begin{equation}\label{eq16}
S = 2 \pi k_B \left(\left|\frac{x}{l_p}\right|+ \left|\frac{l_p}{2x}\right| \right).
\end{equation}
This relation indicates that when $x$ approaching 0, Entropy $S$ becomes infinite. Combining eq.(\ref{eq14}), it is found that
$\delta S$  increases within same $\delta x$  while entropy force increases correspondingly during the procedure of test particle approaching
holographic screen \cite{f28}. This is different from classical assumption $S=\frac{x}{l_p}2 \pi k_B $, ($x \rightarrow 0, S \rightarrow 0$),
that is a new phenomenon introduced by uncertainty relation. Later discussion reveals there is a need to introducing a minimal length
$l_p$ for keeping entropy $S$ in eq.(\ref{eq16}) finite within the interval of one $l_p$ length.\\
\indent The interpretation of the quantum mechanical uncertainty as a form of entropy is perhaps not that strange considering that entropy is
practically equivalent to lack of information regarding an object. In this perspective, once the quantum correction has been introduced, it is
indeed natural to consider the modified wave function of the particle instead of the origin solution of Klein-Gordon equation. Considering
uncertainty of entropy we reach eq.(\ref{eq17}) by comparing (\ref{eq11}) and (\ref{eq16})
\begin{equation}\label{eq17}
|\psi(x)| =Ae^{-\pi \left(\left|\frac{x}{l_p}\right| + \left|\frac{l_p}{2x}\right |\right)}.
\end{equation}
This relation (\ref{eq17})  is our modified wave function of the particle caused by uncertainty principle without phase factor. Indeed, by inspecting
(\ref{eq17}), one can see the main difference between the wave function of a classical particle and the quantum particle up to a factor of reciprocal term. The classical limit is obtained by taking $\delta p =0$ in the above estimate. In original case of eq.(\ref{eq8}), when $x \rightarrow 0$,wave
function $\psi$ collapses to the finite normalization constant $A$, while the corrected wave function collapses to 0 at $x \rightarrow 0$,
and both wave functions collapse to 0 at $x \rightarrow \infty$.\\
\indent We found through numerical calculation that the original wave function reaches maximum at $x=0$ while the corrected one is at
$x=0.707 l_p$. According to the interpretation of classic quantum mechanics, that $|\psi|^2$ means probability, we will certainly receive the particle on the holographic screen if uncertainty relations are not considered. However, when it works we only receive the particle at $x=0.707 l_p$
far from holographic screen.\\
\indent Wave function on the holographic surface collapse to 0 and the entropy taking infinite before both are mostly incredible things. But when we introduce the Planck length $l_p$, these seemingly absurd puzzles become reasonable from the aspect of measurement.\\
\indent Because $l_p$ is the minimum unit of length can be measured, holographic screen, as the super-surface, can not be considered without the thickness, but with the thickness $l_p$. $x \rightarrow l_p$ should be meaningful for us instead of $x \rightarrow 0$ now and the puzzle from
eq.(\ref{eq16}) would be solved\\
\begin{equation}\label{eq18}
S=2 \pi k_B \left(\left|\frac{l_p}{l_p}\right| + \left|\frac{l_p}{2 l_p}\right |\right)=3 \pi k_B.
\end{equation}
Entropy in the holographic interface will no longer take the maximum but a limited value.\\
\indent Similarly, based on the above discussion, $0.707l_p > l_p$, the test particle is not regard at $x=0.707l_p$ far from holographic screen.
Because $l_p$ is the meaningful smallest scale, in other words, it can already be considered on the holographic screen now.
Then the amplitude of the original wave function eq.(\ref{eq8}) should be modified
\begin{equation}\label{eq19}
probability=\int^{l_p} _0 |\psi(x)|^2 dx=0.766.
\end{equation}
\indent This integral as the corrected probability of the particle for measurement would be different from the original one on holographic screen. Furthermore this probability is 0.766 instead of 0 and the puzzle has been solved.\\
\indent We have discussed the physics meaning on the corrected wave function. Considering the uncertainty relation, we illustrated a more accurate procedure of test particle approaching the holographic screen. Our quantum approach to entropy stress the information in a physical system is directly associated with its quantum mechanical entropy defined by the equality of the partition function.

\section{THE SPEED OF INFORMATION TRANSFER}\label{SecD}
\indent In this section, according to the information transfer between test particle and holographic screen in entropic gravity respecting both the uncertainty principle and causality, the speed of information transfer might be derived.\\
\indent The formulation of the entropy mechanism suggests that there is an inherent uncertainty $\triangle x$ in the location of the test mass $m$
relative to the holographic screen. This leads one to suggest that, when the position uncertainty of $m$ is $\triangle x$, there is a statistical fluctuation of the screen's entropy $\triangle S$, and hence an uncertainty in its energy $\triangle E \sim T\triangle S$ that must abide by quantum mechanical
considerations.\\
\indent Here we may re-express the entropy expression (\ref{eq7}) on a differential formulation:
\begin{equation}\label{eq20}
d S = -2 k_B d (ln|\psi|) = -2 k_B |\psi|^{-1} d |\psi|.
\end{equation}
Because an observer in an accelerated frame has the Unruh temperature \cite{f29}:
\begin{equation}\label{eq21}
T=\frac{\hbar a}{2 \pi k_B c}.
\end{equation}
Using (\ref{eq20}) and (\ref{eq21}) we arrive at a change in energy for a particle:
\begin{equation}\label{eq22}
\triangle E = \int dE = \int T dS = -\frac{\hbar}{\pi c} \int \frac{a}{|\psi|} d|\psi|.
\end{equation}
\indent Furthermore, the uncertainty principle further suggests that the fluctuation in the energy of the screen is
constrained to occur during the interval:
\begin{equation}\label{eq23}
\triangle E \triangle t \geqslant \frac{\hbar}{2},
\end{equation}
Once the particle is within a Compton wavelength of the screen, using eq.(\ref{eq23}) the speed of information transfer is:
\begin{equation}\label{eq24}
v \sim \frac{\triangle x}{\triangle t}  \leqslant \frac{2\triangle x \triangle E}{\hbar}.
\end{equation}
Inserting (\ref{eq22}) in above relation we reach:
\begin{equation}\label{eq25}
v \leqslant -\frac{2\lambda\hbar}{\pi c} \int \frac{a}{|\psi|} d|\psi| \leqslant c,
\end{equation}
So
\begin{equation}\label{eq26}
\int \frac{a}{|\psi|} d|\psi| \leqslant -\frac{\pi c^2}{2\lambda\hbar}.
\end{equation}
Here $a$ as gravity is negative. By now we have obtained the speed of information transfer described by the wave function. Furthermore we restrict the value of wave function indirectly and impose a strict causality relation by demanding this upper bound.

\section{Conclusion}\label{SecE}
\indent The nature of the assumption of entropy force, as spoken of by Verlinde \cite{f1}, is addressed in this paper. J. Munkhammar proposed a main uncertainty which build the connection between the quantum entropy arising in quantum mechanics and the thermodynamical entropy \cite{f19}. 
In defence of this assumption, the entropy in quantum mechanics has been accounted further in our discussion 
in the complete thermodynamical-entropy theory of physics . Furthermore, we introduce the model of the uncertainty of entropy \cite{f28}and 
modified the origin wave function of the test particle. The corrected one acting on the quantum particle is subject to the uncertainty principle.
Considering the uncertainty relation, we illustrated a more accurate procedure of test particle approaching the holographic screen. 
The corrected probability of the particle for measurement is different from the original one on holographic screen.
At the last part of this paper we discuss the speed of information transfer described by wave function and impose a strict causality relation by demanding its upper bound. The generality of entropy force is stressed in this paper, which follows from the clearness of analysis. 
Since our arguments rely on the quantum nature of the test particle, it would be interesting to discuss the alternatives to the equipartition theorem which, in general, are model dependent.\\
\indent There are many open problems remaining as this is a theory in progress. The study of multiple particle situations in the quantum mechanical approach should be interesting. A relativistic approach to quantum entropic gravity also needs to be established and investigated. \\

\newpage
\begin{figure}
\begin{center}
\includegraphics[height=6.5cm]{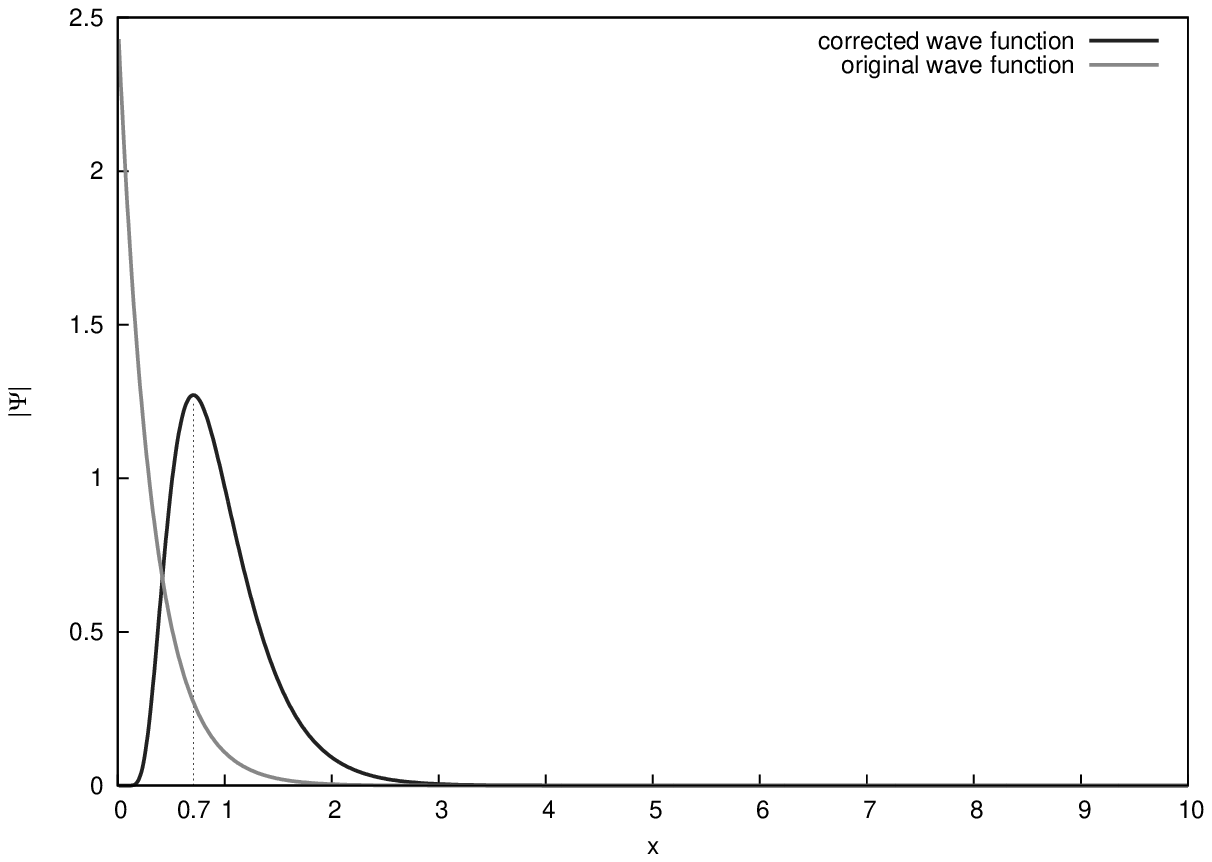}    
\caption{The original wave function and the corrected wave function have been shown in this figure. 
The original wave function reaches maximum at $x=0$ while the corrected one is at $x=0.707 l_p$.} \label{fig:bifurcation}
\end{center}
\end{figure}
\newpage
\begin{figure}
\begin{center}
\includegraphics[height=6.5cm]{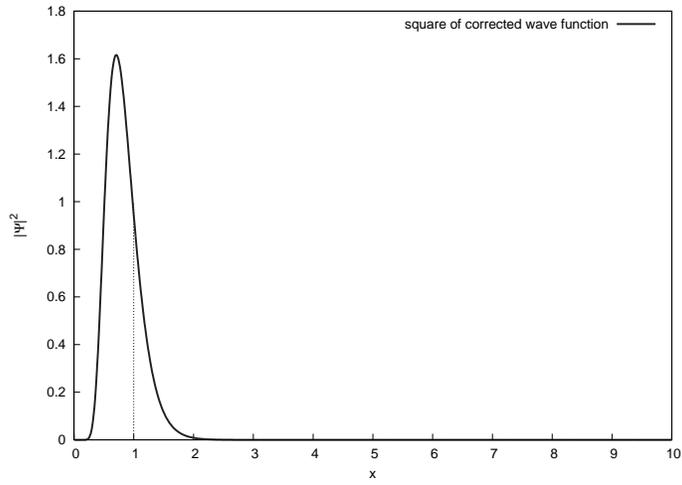}    
\caption{The corrected probability of the particle received on the holographic screen for measurement } \label{fig:bifurcation}
\end{center}
\end{figure}

\end{document}